\documentclass[structabstract]{aa} 

\usepackage{graphics}
\usepackage{graphicx}
\usepackage{amsmath}
\usepackage{amssymb}
\usepackage{listings}
\usepackage{color}
\usepackage{natbib}
\usepackage{txfonts}
\usepackage[T1]{fontenc}
\usepackage[latin1]{inputenc}

\bibpunct{(}{)}{;}{a}{}{,} 

\author{Daniel Johansson\inst{1} \and Haukur Sigurdarson\inst{1}
\and Cathy Horellou\inst{1}}

\institute{Onsala Space Observatory, Department of Earth and Space
  Sciences, Chalmers University of Technology, SE-439 92 Onsala,
  Sweden }

\title{A LABOCA survey of submillimeter galaxies behind galaxy clusters}

\abstract {Submillimeter galaxies are a population of dusty
  star-forming galaxies at high redshift. Measuring their properties
  will help relate them to other types of galaxies, both at high and
  low redshift. This is needed in order to understand the formation
  and evolution of galaxies.  }{The aim is use gravitational lensing
  by galaxy clusters to probe the faint and abundant submillimeter
  galaxy population down to a lower flux density level than what can
  be achieved in blank-field observations.}{We use the LABOCA
  bolometer camera on the APEX telescope to observe five cluster of
  galaxies at a wavelength of 870 $\mu$m. The final maps have an
  angular resolution of 27.5\arcsec and a point source noise level of
  1.2--2.2 mJy. We model the mass distribution in the clusters as
  superpositions of spherical NFW halos and derive magnification maps
  that we use to calculate intrinsic flux densities as well as
  area-weighted number counts. We also use the positions of Spitzer
  MIPS 24~$\mu \mathrm{m}$~sources in four of the fields for a
  stacking analysis.}{We detected 37 submm sources, out of which 14
  have not been previously reported. One source has a sub-mJy
  intrinsic flux density. The derived number counts are consistent
  with previous results, after correction for gravitational
  magnification and completeness levels. The stacking analysis reveals
  an intrinsic 870~$\mu \mathrm{m}$~signal of $390\pm 27$~$\mu$Jy at
  $14.5\sigma$ significance. We study the $S_{\mathrm{24 \, {\mu}m}}$
  -- {$S_{\mathrm{870 \, {\mu}m}}$}~relation by stacking on subsamples
  of the 24~$\mu \mathrm{m}$~sources and find a linear relation at
  $S_{\mathrm{24 \, {\mu}m}}$$<300 \, \mu$Jy, followed by a flattening
  at higher 24~$\mu \mathrm{m}$~flux densities. The signal from the
  significantly detected sources in the maps accounts for 13\% of the
  Extragalactic Background Light discovered by COBE, and the stacked
  signal accounts for 11\%.}  {} \date{\today}


\titlerunning{A LABOCA survey of submillimeter galaxies} 
\authorrunning{Johansson et al.}

\begin{document}

\maketitle

\section{Introduction}
\label{sec:introduction}

Submillimeter galaxies form a population of high-redshift, dusty
star-forming galaxies that are highly obscured in the visible and in
the near-infrared, and have a spectral energy distribution (SED) that
peaks in the submillimeter (submm) waveband (see
e.g. \cite{BlainSmail:2002aa} for a review).  Most recent searches for
submm galaxies have been based on surveys of blank sky with no known
large-scale structure along the line-of-sight.  These surveys have
exploited large-format sensitive bolometer arrays (e.g. SCUBA on the
James Clerk Maxwell Telescope (JCMT), \citealt{CoppinChapin:2006aa};
LABOCA~on the Atacama Pathfinder EXperiment (APEX),
\citealt{WeisKovacs:2009ab}; AzTEC on ASTE:
\citealt{AustermannDunlop:2010aa,ScottYun:2010ab}; the South Pole
Telescope (SPT): \citealt{VieiraCrawford:2010aa}; MAMBO on the IRAM
30~m telescope: \citealt{GreveIvison:2004aa}, 
\citealt{BertoldiCarilli:2007aa}).  Those maps cover large areas at a
nearly uniform noise level, leading to a simple selection function
with a constant completeness across the field. The observations showed
that source number counts increase steeply with decreasing flux
density at mm and submm wavelengths
\cite[e.g.][]{WeisKovacs:2009ab,PatanchonAde:2009ab}. In order to
probe the faint (below a few mJy) population of submm galaxies,
several authors have taken advantage of the gravitational
magnification induced by massive clusters of galaxies
(e.g. \citealt{SmailIvison:1997aa,SmailIvison:2002aa,ChapmanSmail:2002aa,Knudsenvan-der-Werf:2005aa,KnudsenBarnard:2006aa,Knudsenvan-der-Werf:2008aa,JohanssonHorellou:2010ab,WardlowSmail:2010aa,
  RexAde:2009aa,EgamiRex:2010aa}).  A large magnification, produced
for example when a source lies close to a critical line of the lens,
may make it possible to detect a source with an intrinsic flux density
much lower than the formal root mean square of the noise of the
observation. This is the only method of detecting such dim sources
directly.  Cluster field observations have sensitivities that vary
across the map, as magnified sources are ``lifted'' above the
detection limit, and the selection functions are therefore more
complicated.  The most comprehensive study to date of submm galaxies
behind lensing clusters is that of \cite{Knudsenvan-der-Werf:2008aa},
who analyzed SCUBA data from 12 galaxy clusters and one blank field,
resulting in an effective surveyed area of 71.5 arcmin$^2$ on the sky,
but an area in the source plane almost twice as small. Seven sources
with sub-mJy fluxes were detected.

The sources revealed by gravitational lensing are prime targets for
observations across the electromagnetic
spectrum. \cite{SwinbankSmail:2010aa} discovered a very bright submm
source, situated at $z=2.33$, with flux density {$S_{\mathrm{870 \,
      {\mu}m}}$}$\sim$106~mJy, and molecular line observations showed
that the amount of molecular gas is similar to that in local
ultra-luminous infra-red galaxies (ULIRGs,
\citealt{DanielsonSwinbank:2010aa}).  The $z\sim 3.4$ submm source
studied by \cite{IkarashiKohno:2010aa} and discovered through use of
AzTEC at 1.1~mm \citep{WilsonAustermann:2008aa} has a 880~$\mu
\mathrm{m}$~flux density measured by the Submillimeter Array (SMA) of
$\sim 73$~mJy and seems to be a ULIRG as well.  On the other hand, the
$z=2.79$ galaxy behind the Bullet Cluster
\citep{GonzalezPapovich:2010aa}, with a flux density of about 48~mJy
at 870~$\mu \mathrm{m}$, is more representative of the normal galaxy
population with an intrinsic far-infrared luminosity of a few times
10$^{11} L_\odot$
\citep{2008MNRAS.390.1061W,GonzalezClowe:2009aa,JohanssonHorellou:2010ab}. Large
surveys in the mm (SPT) and the far-infrared (Herschel) are
discovering bright lensed submm galaxies
\citep{VieiraCrawford:2010aa,NegrelloHopwood:2010aa}.

Another way to probe the faint part of the submm galaxy population is
to perform a stacking analysis using known positions obtained from
complementary observations at another wavelength.
\cite{2006A&A...451..417D} 
used the positions of sources detected with Spitzer Space Telescope at
24~$\mu$m to measure the contribution of those sources to the 70 and
160~$\mu$m far-infrared background, gaining up to one order of
magnitude in depth. Greve et al. (2010) carried out a stacking
analysis of the LABOCA submm map of the Extended Chandra Deep Field
(ECDF) using a large sample of near-infrared detected galaxies.

\begin{table*}[t!]
  \centering
  \caption{Observed cluster fields.   }
  \begin{tabular}[t]{l l l c c c c }
    \hline
    \hline
    Target & $\alpha$\tablefootmark{a} [J2000] & $\delta$\tablefootmark{a}
    [J2000] & $z$ & rms\tablefootmark{b} & $\Omega$\tablefootmark{c} \\
    & [h~~~~m~~~~s] & [\degr~~~~\arcmin~~~~\arcsec] & &
    [$\mathrm{mJy\, beam}^{-1}$]  & [$\mathrm{arcmin}^2$]\\
    \hline
    {Abell~2163} & 16 15 45.1 & $-$06 08 31 & 0.203 & 2.2 & 150 \\
    Bullet Cluster\tablefootmark{1}  & 06 58 29.2 & $-$55 56 45 & 0.296
    & 1.2 & 220 \\
    Abell~2744\tablefootmark{2} & 00 14 15.0 & $-$30 22 60 & 0.308  &
    1.5 &  220 \\
    AC~114\tablefootmark{3} & 22 58 52.3 & $-$34 46 55 & 0.312 & 1.2 & 130 \\
    MS~1054-03\tablefootmark{4}  & 10 57 00.2 & $-$03 37 27 & 0.823 &
    1.6 & 200 
    \\
    \hline
  \end{tabular}
  \tablefoot{\tablefoottext{a} Central coordinates
    of the $2\times 2$ raster square scanning pattern; these positions
    differ slightly from the central X-ray
    positions. 
    \tablefoottext{b} The noise level measured in the central
    10 arcminutes of each map, as described in
    Sect.~\ref{sec:survey-completeness}. \tablefoottext{c} Extent 
    of the LABOCA~maps. \\ \tablefoottext{1}
    Alternative name 1E~0657-56. Project's observing identification (Obs. ID):
    O-079.F-9304A-2007, E-380.A-3036A-2007. \\ \tablefoottext{2}
    Alternative name AC 118. Obs. ID
    O-081.F-9319A-2008. \\ \tablefoottext{3} Alternative name Abell~S~1077, Obs. ID E-081.A-0451A-2008,
    E-078.F-9032A-2007. \\  \tablefoottext{4} Obs. ID O-083.F-9300A-2009}
  \label{tab:targ}
\end{table*}

In this paper, we extend the analysis of submm sources behind the
Bullet Cluster~recently presented in \cite{JohanssonHorellou:2010ab}
by four additional galaxy cluster fields observed with the
LABOCA~receiver on the APEX telescope. The deep observations allow us
to detect submm galaxies with observed flux densities above $\sim
4.5$~mJy, while the gravitational magnification reveals galaxies with
intrinsic sub-mJy flux densities. We derive the magnification of the
foreground clusters by using the lens equation for clusters modeled as
a superposition of Navarro, Frenk and White (NFW, 1997) mass density
profiles whose parameters are inferred from published papers on the
selected clusters.  From the magnification maps, we calculate
intrinsic flux densities and derive submm number counts for the entire
survey. We carry out a stacking analysis on 24~$\mu
\mathrm{m}$~detected sources in the fields to probe the correlation
between submm and mid-infrared emission and detect stacked 870~$\mu
\mathrm{m}$~observed flux densities of {$S_{\mathrm{870 \,
      {\mu}m}}$}~$<800$~$\mu$Jy for sources that are undetected
individually in the maps.

This paper is organized as follows: in Sect.~\ref{sec:observations} we
describe the submm observations and data reduction and the Spitzer
MIPS 24~$\mu \mathrm{m}$~archival data; in Sect.~\ref{sec:results} we present
the resulting maps. In Sect.~\ref{sec:cluster-lens-models} we discuss
the lensing models and the number counts and in
Sect.~\ref{sec:stacking-analysis} we present a stacking
analysis. Section~\ref{sec:contribution-ebl} discusses the
contribution of our submm signals to the Extragalactic Background
Light discovered by COBE. The results are summarized in
Sect.~\ref{sec:conclusions}.

Throughout the paper, we adopt the following cosmological parameters:
a Hubble constant $H_0 = 71$~km~s$^{-1}$~Mpc$^{-1}$, a matter density
parameter $\Omega_0 = 0.27$, and a dark energy density parameter
$\Omega_{\Lambda0} = 0.73$. The redshift $z=0.3$ where three of our
clusters reside corresponds to an angular-diameter distance of 911~Mpc
and a scale of 4.42~kpc/arcsec. $z=2.2$, the median redshift of known
submm galaxies, corresponds to an angular-diameter distance of
1728~Mpc and a scale of 8.38~kpc/arcsec\footnote{We used Ned Wright's
  cosmology calculator \citep{Wright:2006aa} available at
  \texttt{http://www.astro.ucla.edu/{\textasciitilde}wright/cosmocalc.html}.}.

\section{Observations and data reduction}
\label{sec:observations}

We have gathered data from galaxy cluster fields observed with the
LABOCA~bolometer camera on the APEX\footnote{This publication is based
  on data acquired with the Atacama Pathfinder EXperiment (APEX). APEX
  is a collaboration between the Max-Planck-Institut f\"ur
  Radioastronomie, the European Southern Observatory, and the Onsala
  Space Observatory} telescope in Chile \citep{Gusten:2006ak}. The
five clusters clusters are merging systems, and their high masses
yield areas of large gravitational magnification, which increases the
possibility of finding very intrinsically dim submillimeter sources
lensed by the cluster. Three of the cluster field observations are
from our own observing programs, while the AC~114~data (Principal
Investigators S. Chapman and F. Boone) were downloaded from the ESO
archive and the Abell~2163~data were provided by the
PI. M. Nord. Detailed information about the cluster fields, including
integration time and noise levels of the final maps, is given in
Table~\ref{tab:targ}.

Ground-based submm observations suffer from the fact that the Earth's
atmosphere is by far brighter than the astronomical sources. The
changing temperature of the atmosphere further complicates the data
reduction. The relatively small field-of-view of
LABOCA~($11.4\arcmin$) limits the influence of the spatial temperature
structure of the atmosphere on the measurement.

LABOCA~observes in total power scanning mode, where the telescope
scans the sky in a pattern that is designed to facilitate the
retrieval of the astronomical signal and the removal of the
atmospheric signal.  The scanning pattern that was used for our
observations is an outwards winding spiral which is repeated at four
raster points. At a given time during the scan, the atmospheric signal
is correlated across the entire array and we can model and remove
it. The faint astronomical signal is not correlated, unless it is
distributed on scales comparable to the field-of-view of the bolometer
camera. The {\tt Minicrush}~software \citep{Kovacs:2008ab}, that we
use to reduce the data, utilizes this approach when removing the
correlated atmospheric noise.

Several types of calibration data are taken during the
observations. Absolute flux calibration is determined from
observations of the primary calibrators: Neptune, Uranus and
Mars. When no primary calibrator is available, secondary calibrators,
which are well studied objects for which the flux ratios to the
primary calibrators are known, can be used. Measuring the calibrators
also gives a measure of the opacity of the atmosphere. The opacity is
also measured by performing \emph{skydips}, which are fast scans that
measure the sky temperature as a function of elevation at constant
azimuthal angle. These scans are performed every 2-3 hours. The
calibration of LABOCA data is described in detail by
\cite{SiringoKreysa:2009aa}. The telescope pointing was checked
regularly with scans on nearby bright sources and was found to be
stable within 3\arcsec~(rms). The angular resolution (FWHM) of LABOCA
on APEX is 19.5\arcsec.

\subsection{Data reduction}
\label{sec:data_reduction}

The data were reduced using the {\tt Minicrush}~software
\citep{Kovacs:2008ab}, similarly to the procedure described in
\cite{JohanssonHorellou:2010ab}. We summarize the steps here. The data
are organized in MBFITS-files, where data from each bolometer as a
function of time are saved in a so-called timestream. Each scan, and
thereby each MBFITS-file, contains the timestream data of each
bolometer. {\tt Minicrush}~attempts to remove the correlated noise by
temporarily regarding it as a signal, and fitting a model to all the
timestreams at the same time. This model is then removed from each
timestream, and the result is a cleaner signal, with less correlated
noise. This procedure is repeated a number of times (for
LABOCA~usually six to eight times) until the resulting signal is
"white", that is that most of the $1/f$-type noise has been removed.

An advantage of this method for removing correlated noise is that the
gains of each individual bolometer can be estimated during the
process. Another method for determining the gains is to observe a
bright calibration source and scan it to produce a fully sampled map
with each bolometer. The information about the gains is used to
flatfield the data. It can also be used to flag and remove suspicious
bolometer channels from the reduction. A channel with almost no
optical response will appear to have very low noise level, but
searching for and flagging channels with low gain would find and
remove that channel from the reduction. The pipeline also flags spikes
and glitches in the bolometer channels.

We used the option \texttt{'--deep'} in {\tt Minicrush}. This turns on
the most aggressive filtering and is useful when searching for
point-like sources. Extended structures, such as the Sunyaev-Zeldovich
increment from the clusters, are filtered out.

\subsection{Making maps}
\label{sec:making-map}

When the pipeline has removed the correlated noise from the atmosphere
and from the instrument, flagged any optically dead channels and any
bad pixels (for example hit by cosmic rays), the timestreams should be
``white'', i.e. free from $1/f$-type noise. The astronomical signal is
typically too weak to be seen in the timestreams, and maps from
individual scans have to be produced and co-added to reduce the
noise. The maps are made by using the scanning pattern of the
telescope and map each bolometer position onto a grid of points; when
a bolometer has ``seen'' a certain pixel on the map, its flux is
deposited there. Since several bolometers have seen the same portion of
the sky, the final flux value in one map pixel is an average of the
flux of the bolometers that observed that part of the sky, weighted by
the variance of the individual bolometers. A noisy bolometer thus
contributes less to the flux density value in a single pixel in the
map than a less noisy one.

Together with the flux density map (the ``signal'' map), a \emph{noise
  map} is created. The coadded values of the time-stream weights are
used to create the noise map. A signal-to- noise map, which is the
signal map divided by the noise map, is also appended to the FITS-file.

\subsection{Spitzer MIPS data reduction}
\label{sec:spitzer-mips-data}

In Sect.~\ref{sec:stacking-analysis} we discuss a stacking analysis in
the LABOCA~maps on Spitzer MIPS \citep{RiekeYoung:2004aa}
24~$\mu \mathrm{m}$~source positions. We describe here the MIPS data reduction
and source extraction procedure. We follow the general recommendation
not to use the pipeline-processed MIPS mosaics, but to re-process the
data, because persistent images can create artefacts on the
mosaics. Since we use only the 24~$\mu \mathrm{m}$~MIPS data, ``MIPS'' will
refer to that band in the remainder of this paper.

Four of the clusters in our survey have been observed in the
24~$\mu \mathrm{m}$~MIPS band. We used all the available 24~$\mu \mathrm{m}$~data for
each cluster field, as summarized in Table~\ref{tab:mips}. We started
by downloading the basic calibrated data from the Spitzer science
archive. For each Astronomical Observation Request (AOR), we created a
flatfield frame using the script
``\texttt{flatfield\_24\_ediscs.nl}''. That frame was then used to
correct for any persistent problem in the data. We then used
\texttt{Mopex} to do overlap correction on all the data for each
target. The overlap-corrected data were then mosaiced using
\texttt{Mopex}. We used the default values for all the steps in the
pipeline.

Source extraction from the mosaics was performed with the
\texttt{Apex} tool. We did Point Response Function (PRF) fitting and
aperture photometry to detect significant MIPS sources. The number of
detected sources per field is listed in Table~\ref{tab:mips}, together
with the MIPS coverage across the LABOCA~map, the median noise level for
point sources and the 24~$\mu \mathrm{m}$~source number density. For the Bullet
Cluster and MS 1054-03, the MIPS maps cover almost the entire LABOCA
field, but for AC 114 and Abell 2744 the MIPS map are significantly
smaller.

\begin{table}[h!]
  \caption{Summary of archival MIPS data used in this study.}
  \centering
  \begin{tabular}{l c c c c}
    \hline
    \hline
    Cluster name & $\Omega$\tablefootmark{a} & $\sigma$\tablefootmark{b} &
    $N_{\mathrm{s}}$\tablefootmark{c} & $\Sigma$ \\
    & \tiny{[$\mathrm{arcmin}^2$]} & \tiny{[$\mu$Jy]} & & \tiny{[$10^{3}$~deg$^{-2}$]} \\
    \hline
    \vspace{0.5ex}
    Bullet Cluster\tablefootmark{1} & $20\arcmin \times 21\arcmin$ & -- & 325
    & 2.8 \\
    \vspace{0.5ex}
    Abell~2744\tablefootmark{2} & $6\arcmin \times 11\arcmin$ & 13.8 & 193  & 10.5 \\
    \vspace{0.5ex}
    AC~114\tablefootmark{3} & $8\arcmin \times 11\arcmin$ & 10.0 & 208 & 8.5 \\
    \vspace{0.5ex}
    MS~1054-03\tablefootmark{4} & $17\arcmin \times 19\arcmin$ & 20.3 & 552 &
    6.2 \\
    \hline
  \end{tabular}
  \tablefoot{\tablefoottext{a} Angular coverage across the LABOCA~field.
    \tablefoottext{b} Median $1\sigma$ noise level for extracted
    24~$\mu \mathrm{m}$~sources in the part of the map covered by
    LABOCA. \tablefoottext{c}Number of sources in the area
    reported in column 2. \\ \emph{Program identification numbers} (PIDs) \\ \tablefoottext{1} 40137, 40593 \\
    \tablefoottext{2} 83, 3644 \\  \tablefoottext{3} 83, 50096 \\
    \tablefoottext{4} 20740, 83, 3644, 50726 }
 \label{tab:mips}
\end{table}

\section{Results}
\label{sec:results}

\begin{figure*}[h!]
  \includegraphics[width=16cm]{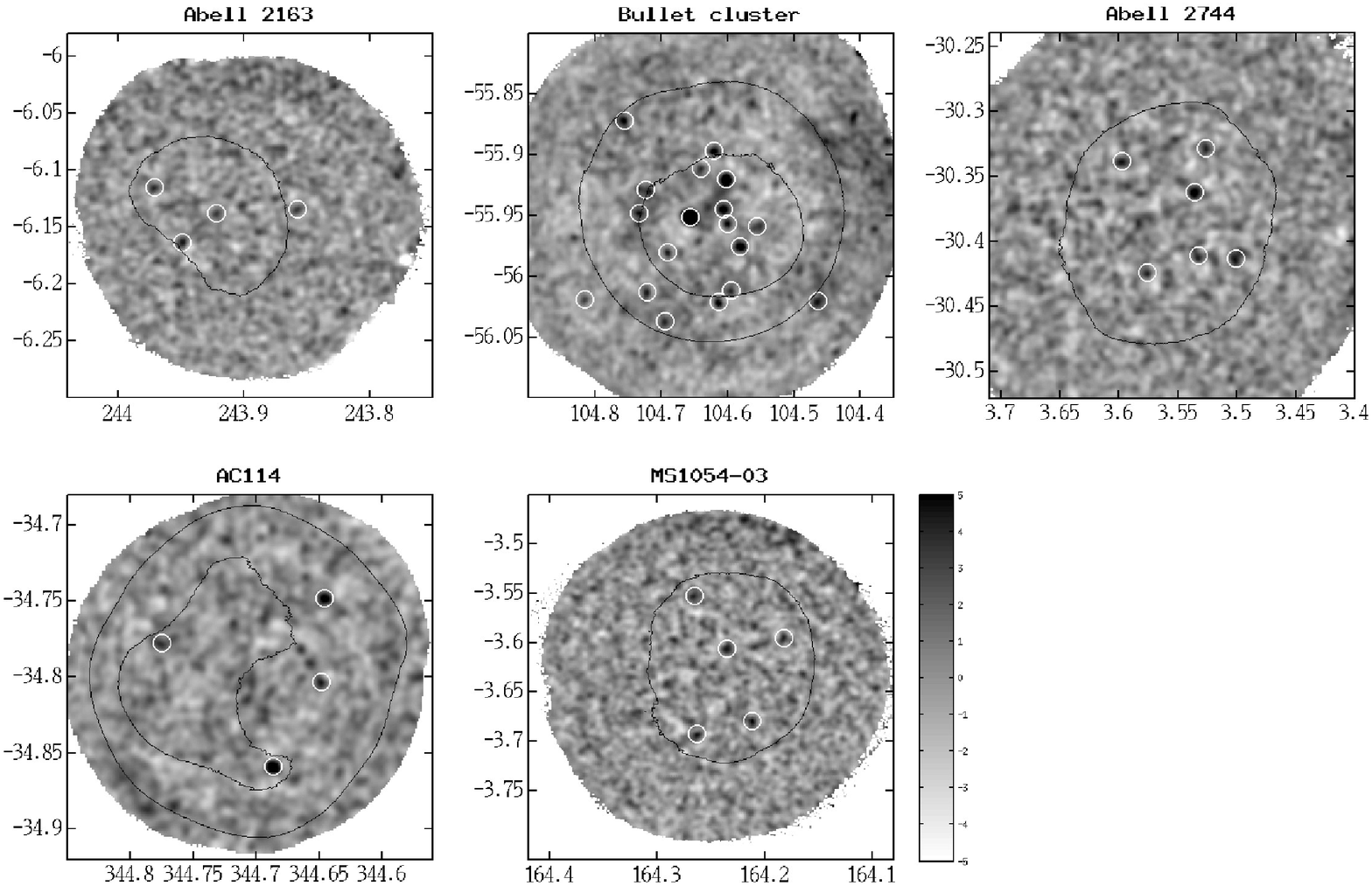}
  \caption{Signal-to-noise maps of the five cluster fields. White
    circles represent the significant sources in the map and black
    contours show the noise maps of each cluster map at levels of 2, 4
    and 8 $\mathrm{mJy\, beam}^{-1}$. The signal-to-noise representation causes the
    appearance of the increasing noise towards the edge of each map to
    be suppressed.}
  \label{fig:snmaps}
\end{figure*}

Figure~\ref{fig:snmaps} shows the final reduced signal-to-noise maps
of the cluster fields. They have been smoothed with a Gaussian of the
size of the beam (19.5\arcsec) to a final resolution of 27.5\arcsec,
and emission on scales larger than 100\arcsec~has been filtered
out. Contours of the noise maps are overlaid on the signal-to-noise
maps.

In the remainder of this section we discuss the source extraction
process, and describe the source catalog and the Monte Carlo
simulations that we use to characterize the noise level and
completeness in each field. We also discuss the method we used for
flux deboosting and how we estimate the number of spurious detections.

\subsection{Source extraction}
\label{sec:source-extraction}

\begin{table*}[t!]
  \caption{Flux densities and magnification
    factors for submm sources detected in the survey.}
  \centering
  \tiny{
  \begin{tabular}{l c c c l c}    
    \hline
    \hline
    Submm source name & {$S_{\mathrm{870 \, {\mu}m}}$}\tablefootmark{a} & $S_{\mathrm{deboosted}}$\tablefootmark{b} & $S_{\mathrm{demag}}$
    & $\mu$ & S/N\tablefootmark{c} \\
    & [mJy] & [mJy] & [mJy]  & & \\
    \hline
    \emph{Abell~2163} & & & \\ 
    SMM~J161525.8-060803 &$  7.8 \pm 2.2$ & $  4.3 \pm   3.5$ &   3.8 &  1.12  &    3.5 \\ 
    SMM~J161541.2-060817\tablefootmark{2} &$  4.9 \pm 1.8$ &  --- &   --- &  1.52  &    2.8 \\ 
    SMM~J161547.7-060948 &$  8.9 \pm 2.0$ & $  6.4 \pm   2.3$ &   3.1 &  2.04  &    4.4 \\ 
    \vspace{1ex} 
    SMM~J161553.1-060655 &$  6.7 \pm 1.8$ & $  4.5 \pm   2.3$ &   2.9 &  1.56  &    3.8 \\ 
    \emph{1E~0657-56} & & & \\ 
    SMM~J065751.4-560112 &$ 13.5 \pm 2.7$ & $  9.7 \pm   3.0$ &   8.9 &  1.09  &    5.0 \\ 
    SMM~J065813.4-555732 &$  4.9 \pm 1.0$ & $  4.2 \pm   1.0$ &   2.8 &  1.54  &    5.1 \\ 
    SMM~J065819.4-555830 &$  8.2 \pm 0.9$ & $  7.7 \pm   0.9$ &   5.0 &  1.55  &    8.7 \\ 
    SMM~J065822.9-560041 &$  4.8 \pm 1.2$ & $  3.8 \pm   1.3$ &   3.0 &  1.28  &    4.2 \\ 
    SMM~J065824.0-555723 &$  5.3 \pm 0.9$ & $  4.7 \pm   1.0$ &   2.2 &  2.10  &    5.7 \\ 
    SMM~J065824.5-555512 &$ 15.1 \pm 1.0$ & $ 14.7 \pm   1.0$ &   9.2 &  1.59  &   14.9 \\ 
    SMM~J065825.5-555640 &$  6.9 \pm 0.9$ & $  6.4 \pm   1.0$ &   2.9 &  2.21  &    7.3 \\ 
    SMM~J065827.3-560116 &$  9.0 \pm 1.3$ & $  8.0 \pm   1.3$ &   6.4 &  1.25  &    7.0 \\ 
    SMM~J065828.9-555349 &$  9.3 \pm 1.2$ & $  8.6 \pm   1.2$ &   6.4 &  1.35  &    7.9 \\ 
    SMM~J065833.7-555441 &$  4.6 \pm 1.1$ & $  3.6 \pm   1.2$ &   2.3 &  1.58  &    4.1 \\ 
    SMM~J065837.6-555705\tablefootmark{1} &$ 48.6 \pm 1.3$ & $ 48.0 \pm   1.3$ &   0.6 &  75.0\tablefootmark{3}  &   36.7 \\
    SMM~J065845.6-555848 &$  6.2 \pm 1.1$ & $  5.5 \pm   1.1$ &   3.1 &  1.79  &    5.9 \\ 
    SMM~J065846.6-560212 &$  7.2 \pm 1.9$ & $  4.6 \pm   2.5$ &   3.9 &  1.18  &    3.8 \\ 
    SMM~J065853.2-560046 &$  7.8 \pm 1.5$ & $  6.4 \pm   1.6$ &   5.1 &  1.25  &    5.2 \\ 
    SMM~J065853.7-555543 &$  5.5 \pm 1.2$ & $  4.5 \pm   1.2$ &   2.9 &  1.55  &    4.7 \\ 
    SMM~J065856.0-555652 &$  5.4 \pm 1.2$ & $  4.4 \pm   1.3$ &   3.0 &  1.50  &    4.6 \\ 
    SMM~J065901.3-555218 &$ 11.9 \pm 2.1$ & $  9.7 \pm   2.1$ &   8.5 &  1.14  &    5.8 \\
    \vspace{1ex} 
    SMM~J065915.6-560108\tablefootmark{2} &$ 23.6 \pm 5.9$ & --- &  --- &  1.11  &    4.0 \\ 
    \emph{Abell~2744} & & & \\ 
    SMM~J001400.2-302447 &$  8.1 \pm 1.5$ & $  6.8 \pm   1.6$ &   6.2 &  1.09  &    5.4 \\ 
    SMM~J001406.3-301942 &$  7.4 \pm 1.4$ & $  6.1 \pm   1.5$ &   5.0 &  1.23  &    5.2 \\ 
    SMM~J001407.7-302439 &$  5.8 \pm 1.2$ & $  4.8 \pm   1.3$ &   4.1 &  1.17  &    4.7 \\ 
    SMM~J001408.6-302142 &$  9.0 \pm 1.3$ & $  8.0 \pm   1.3$ &   5.2 &  1.55  &    6.9 \\ 
    SMM~J001418.3-302525 &$  4.7 \pm 1.3$ & $  3.3 \pm   1.6$ &   2.9 &  1.17  &    3.6 \\ 
    \vspace{1ex} 
    SMM~J001423.4-302018 &$  6.9 \pm 1.5$ & $  5.5 \pm   1.6$ &   3.5 &  1.54  &    4.6 \\ 
    \emph{AC~114} & & & \\ 
    SMM~J225835.0-344453 &$  9.2 \pm 1.4$ & $  8.0 \pm   1.4$ &   7.0 &  1.15  &    6.6 \\ 
    SMM~J225835.7-344812 &$  5.0 \pm 1.5$ & $  3.1 \pm   2.0$ &   2.1 &  1.46  &    3.4 \\ 
    SMM~J225844.7-345131 &$ 11.4 \pm 1.2$ & $ 10.6 \pm   1.2$ &   8.3 &  1.28  &    9.4 \\ 
    \vspace{1ex} 
    SMM~J225905.9-344639 &$  5.1 \pm 1.1$ & $  4.2 \pm   1.2$ &   3.6 &  1.19  &    4.6 \\ 
    \emph{MS~1054-03} & & & \\ 
    SMM~J105643.7-033543 &$  6.8 \pm 1.5$ & $  5.3 \pm   1.7$ &   5.3 &  1.01  &    4.4 \\ 
    SMM~J105650.8-034046 &$  7.9 \pm 1.5$ & $  6.5 \pm   1.6$ &   6.4 &  1.01  &    5.2 \\ 
    SMM~J105656.4-033622\tablefootmark{1} &$  9.8 \pm 1.8$ & $  6.5 \pm   1.3$ &   4.4 &  1.46  &    6.8 \\ 
    SMM~J105703.2-034135 &$  8.4 \pm 1.6$ & $  6.8 \pm   1.7$ &   6.8 &  1.01  &    5.1 \\ 
    SMM~J105703.7-033309 &$  8.3 \pm 1.7$ & $  6.5 \pm   1.8$ &   6.5 &  1.00  &    5.0 \\ 
    \hline
  \end{tabular} }
\tablefoot{ \tablefootmark{a} The flux density is
  obtained by fitting a beam-shaped Gaussian to the LABOCA~source,
  while the
  uncertainty is measured in the jack-knife maps, as discussed in the
  text. \\ \tablefootmark{b} By construction, the 
  deboosted flux densities have asymmetric
  uncertainty intervals, but the difference between the upper and
  lower uncertainties is smaller than the
  number of significant digits reported here. \\ \tablefootmark{c}
  Signal-to-noise ratio of each detection calculated from the observed
  values in column 2. This value may be
  smaller than the threshold for source extraction ($3.5\sigma$) which
  is imposed in the signal-to-noise map. \\
  \tablefoottext{1} Extended sources. \tablefoottext{2}
  Sources for which the posterior flux distribution from the flux
  deboosting algorithm has no local maximum at
  $S>0$~mJy. SMM~J161541.2-060817 is very close to the detection
  threshold, and it is possible that it is not a real
  source. It was excluded from the number counts calculation. SMM~J065915.6-560108 was discussed in
  \cite{JohanssonHorellou:2010ab} and has both AzTEC 1.1~mm and
  Herschel SPIRE counterparts. See also the discussion about
  spurious detections in Sect.~\ref{sec:spurious-detections}.
  \tablefoottext{3} Our derived magnification for this highly
  magnified source is $\mu \sim 41$, which is different from
  the value of 75 which was derived from more detailed modelling by
  \citet{GonzalezClowe:2009aa}, and thus we adopt their value.  }
 \label{tab:slist}
\end{table*}

We impose a detection threshold of $>3.5\sigma$ in the signal-to-noise
maps for source extraction. We exclude sources that lie on the edge of
the maps where the signal-to-noise representation is not accurate.

In each map we extract any source position with significance
$>3.5\sigma$ and fit a circular, two-dimensional Gaussian to the same
position in the signal map. We limit the size of the Gaussian to that
of the beam's FWHM of the images, because submm galaxies are very likely
to be point-like with respect to the
LABOCA~beam. \cite{TacconiNeri:2006aa} found a median source-sizes of
$<0.5\arcsec$, derived from interferometric CO line emission
observations, in a sample of submm galaxies. In two cases we fit
elliptical Gaussians, where we know from previous observations that
the LABOCA~sources are comprised of emission from two or more
galaxies. The two sources are (1) the brightest source in the Bullet
cluster (SMM~J065837.6-555705), which is known to be a blend of two
images of the same $z=2.79$ galaxy \citep[see
e.g.][]{2008MNRAS.390.1061W,JohanssonHorellou:2010ab,GonzalezPapovich:2010aa},
strongly lensed by the cluster potential and an elliptical cluster
member; (2) SMM~J105656.4-033622 in MS~1054-03~which has three SCUBA
850~$\mu \mathrm{m}$~counterparts as reported by
\cite{Knudsenvan-der-Werf:2008aa}, and is discussed further in
Appendix~\ref{sec:comp-with-other}.

We measure the noise level for each source from the Monte Carlo
simulated maps described in the next section. The procedure ensures
that neither confusion noise nor nearby sources contaminate the noise
estimate. 

The final source list is displayed in Table~\ref{tab:slist}. There, we
list, together with positions and measured flux densities of each
source, deboosted flux densities and gravitational magnification
values, which are discussed in the following sections.

\subsection{Survey completeness and depth}
\label{sec:survey-completeness}

Monte Carlo simulations are used to analyze the noise levels of the
observations and to simulate the completeness of the survey. We create
so-called jack-knife maps\footnote{Jack-knifing is a general
  statistical technique used in all fields of science to estimate the
  precision of sample statistics, and it has been used by several
  groups to analyze mm/submm bolometer data.}, which are the
result of coadding all the scanmaps on one target when multiplying
half of the scans with $-1$. This effectively removes any astronomical
emission from the resulting maps, and makes them representations of
the instrumental noise only. By randomizing the positive and negative
scans a large number of different jackknives can be created, all of
which being random realizations of the noise in our observations. For
each cluster field we created 500 jack-knife maps.

We note that the confusion noise in the real maps is effectively
removed from the jack-knifed maps. This implies that the noise level
is underestimated. One can show that at the depth of our maps, the
instrumental noise exceeds the confusion noise. To estimate the
confusion level, i.e. the flux level where a larger integration time
will not decrease the noise level due to the unresolved background
sources, we use the standard estimate that confusion occurs when there
is one source per 30 beams \citep{Condon:1974aa,Hogg:2001aa}. We can
estimate the confusion level from the relations presented by
\cite{Knudsenvan-der-Werf:2008aa}. They used a power law distribution
for the number counts ($N(>S)=N_0 S^{-\alpha}$, with $N_0=13000 \,
\mathrm{deg}^{-2}$ and $\alpha=2.0$) of submm galaxies
\citep{BargerCowie:1999aa,BorysChapman:2003aa}. The confusion noise
level is then $S_{\mathrm{conf}}= \left( 30\Omega_{\mathrm{beam}} N_0
  \mu^{1-\alpha} \right)^{1/\alpha}$ where $\mu$ is the mean
gravitational magnification across the field. Thus, the confusion
level is lowered by the lensing, and for the map FWHMs of
27.5\arcsec~and a mean magnification factor of 1.5 (which is a lower
limit), the confusion level is $<3$~mJy, which is lower than the
faintest detected source in our survey, at 4.6~mJy. The confusion
noise is thus much smaller than the instrumental noise, and can be
safely neglected in the following analysis.

\subsubsection*{Noise levels}
\label{sec:noise-levels}

\begin{figure}[h!]
  \centering
  \includegraphics[width=8cm]{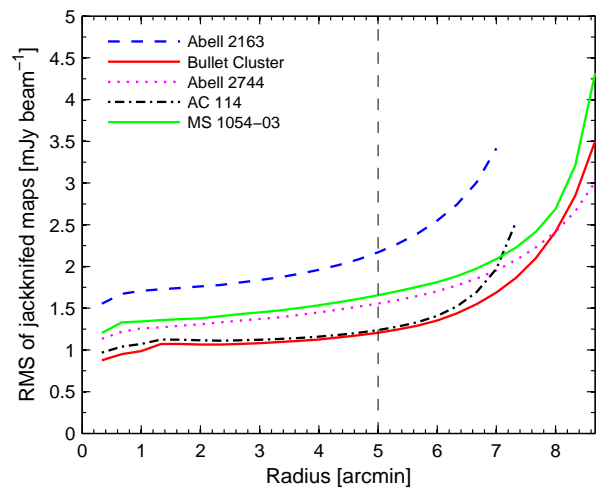}
  \caption{Pixel-to-pixel root-mean-square (rms) as a function of
    radial distance from the map center for the five cluster
    fields, described in Sect.~\ref{sec:completeness}.}
  \label{fig:radrms}
\end{figure}

The jack-knifed maps are used to estimate the noise levels of each map
as a function of angular distance from the center. In a circle of
increasing size we extract all pixels in each jack-knife map, measure
the standard deviation, and then take the average of the values for
all the jack-knife maps. This procedure is repeated for increasing
values of the radial coordinate. The histograms of the distribution of
pixel values in the jack-knife maps are well described by Gaussians,
whose standard deviations are measures of the noise level in each
cluster map.  In Fig.~\ref{fig:radrms} we show the results of that
analysis for the five cluster fields. It can be seen that the noise
level changes very little out to a radius of 5\arcmin. The noise level
reached at that distance is the value that we report for each map in
Table~\ref{tab:targ}. Also, because the noise level is almost constant
in this region, it is the part of the map that we use in the number
counts.

From the jack-knives we also estimate the noise level for each
detected source. Around the position of the submm source (which is not
present in the jack-knife maps) we extract a circular area the size of
the beam, and measure the standard deviation for those pixels in each
jack-knife map. The average value is reported as the uncertainty in
the second column in Table~\ref{tab:slist}.

\subsubsection*{Completeness}
\label{sec:completeness}

We simulate the effects of completeness by inserting artificial
sources (Gaussians of the size of the LABOCA beam) into randomly
chosen jack-knife maps, running the source extraction algorithm, and
then comparing the detected sources to those input. We limit the
angular area to the central 10\arcmin. Similar analysis have
previously been performed by
\cite{BeelenOmont:2008aa,Knudsenvan-der-Werf:2008aa,WeisKovacs:2009ab,JohanssonHorellou:2010ab}.
We simulate sources of flux densities from 1 to 15 mJy, increasing
incrementally by 0.5 mJy, and make 500 simulations per flux density
bin. Although the jackknife maps are realizations of the noise in the
maps, it is possible (and consistent with the underlying Gaussian
statistics) to find fake ``sources'' which are noise peaks. We
therefore include the condition that a detected source should be
situated sufficiently close (within a beam) of the input source.

\begin{figure}[h!] 
 \centering
  \includegraphics[width=8cm]{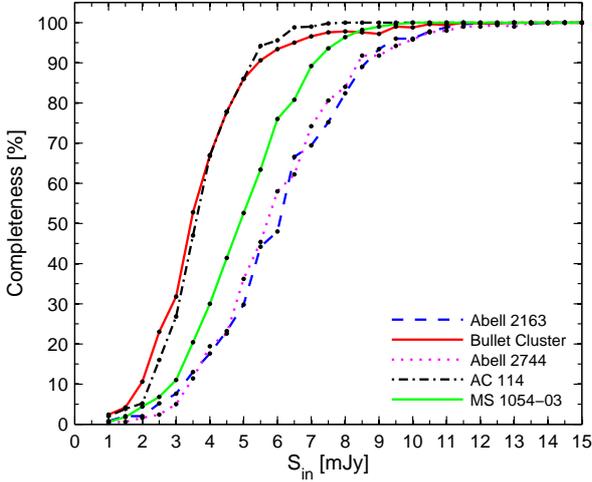}
  \caption{Completeness of the five cluster fields as a function of
    flux density, as described in Sect.~\ref{sec:completeness}.}
  \label{fig:compl}
\end{figure}

The results for the completeness simulations of the five LABOCA maps
are displayed in Fig.~\ref{fig:compl}. The completeness curves follow
the general expected behavior; a noisier map has a lower completeness
value at a certain flux density. From the curves we see that for
example the Bullet Cluster map is $\sim 70\%$ complete at 4.2 mJy (the
$3.5\sigma$ limit for source extraction), while MS~1054-03~at the
corresponding flux density of 5.4~mJy the map is 65\% complete. At a
flux density of 6~mJy the maps are 48\% (Abell~2163), 93\% (Bullet Cluster),
58\% (Abell~2744), 96\% (AC~114) and 77\% (MS~1054-03) complete.

The completeness curves are used to evaluate the submm number
counts. With the $3.5\sigma$ significance limit for source extraction
we must take undetected sources into account when constructing the
number counts, which are discussed in Sect.~\ref{sec:number-counts}.

\subsection{Flux deboosting}
 \label{sec:flux-deboosting}

 The faint population of submm sources acts to boost the flux density
 of the detected sources in the survey. We have used a Bayesian recipe
 \citep{CoppinHalpern:2005aa} to correct for this flux boosting (see
 e.g.~\citealt{ScottAustermann:2008aa,HoggTurner:1998aa}). The
 procedure is described in detail in Appendix A of
 \cite{JohanssonHorellou:2010ab}. A prior flux distribution is
 calculated in a Monte Carlo simulation, where we create sky maps with
 sources distributed in flux according to a Schechter distribution,
 and source positions are drawn randomly, ignoring the effects of
 clustering\footnote{Clustering of faint submm sources have been
   observed in several studies (see
   e.g. \citealt{GreveIvison:2004aa,2006MNRAS.370.1057S,WeisKovacs:2009ab}),
   but at a resolution of 27.5\arcsec~the confusion noise contribution
   from clustering is much smaller than that from the ``normal''
   Poisson distributed confusion noise
   \citep{NegrelloMagliocchetti:2004aa}, and can therefore safely be
   neglected.}. We generate $10^{6}$ simulated maps, and calculate a
 mean flux distribution from the histogram of pixel values for each
 map. This histogram is our prior flux distribution. The prior is
 multiplied with the probability of measuring a flux density $S_m$
 when the intrinsic flux density is $S_i$. This probability is modeled
 as a Gaussian distribution, with the observed flux and noise levels
 in column 2 of Table~\ref{tab:slist} as mean and dispersion. The
 product of the prior flux distribution and the probability to measure
 a flux $S_m$ when the intrinsic flux is $S_i$ is normalized to yield
 the posterior flux distribution\footnote{See the Figure~A1 in
   \cite{JohanssonHorellou:2010ab} for an illustration of these
   operations.}. The deboosted flux density corresponds to the
 $x$-axis value found at the local maximum of the posterior flux
 distribution. These values are listed in Table~\ref{tab:slist}. Due
 to the monotonic decrease of the prior flux distribution, the process
 yields non-symmetric flux density uncertainties, but as noted in
 Table~\ref{tab:slist}, the difference between the upper and lower
 uncertainty is smaller than the number of given digits in the table.

We model the prior distribution by a Schechter function with the
following form 
\begin{equation}
  \label{eq:2}
  \frac{d N}{d S} = N' \left( \frac{S}{S'} \right)^{\alpha + 1} \exp{
    \left(- S/S' \right)},
\end{equation}
\citep{Schechter:1976aa} with parameters from the SHADES survey
\citep{CoppinChapin:2006aa}, that have been scaled from 850~$\mu \mathrm{m}$~to
870~$\mu \mathrm{m}$~using a submm spectral index of 2.7. That yields the
following parameter values: $N' = 1703 \, \mathrm{deg}^{-2} \,
\mathrm{mJy}^{-1}$, $S' = 3.1 \, \mathrm{mJy}$ and $\alpha = -2.0$.

\subsection{Spurious detections}
\label{sec:spurious-detections}

Our adopted criterion for source extraction ($S/N \ge 3.5$)
  means that we are detecting sources close to the noise in the
  maps. It is therefore possible that some of our detections are
  spurious; they might be due to a noise peak boosted by confusion
  noise or instrumental artefacts. It is important to investigate how
  many of the sources in our catalog that might be spurious
  detections. We do that by employing two different techniques:
\begin{enumerate}
\item \emph{The number of negative 3.5$\sigma$ peaks in the maps}:
  We run the source detection algorithm on inverted maps with the same
  3.5$\sigma$ criterion, to estimate the number of spurious
  detections. In the five inverted maps we find four sources,
  indicating that at least four of the sources in our catalog could be
  spurious detections.
\item \emph{The probability that a source has a negative deboosted
    flux density}: the deboosting algorithm gives us the posterior
  flux density for each flux/noise-pair, which is a probability
  distribution for the flux density. For the 37 sources in the survey, we
  find four that have a $\ge 5\%$ probability of having a negative
  flux density, and are possibly spurious. These four sources include
  SMM~J161541.2-060817 and SMM~J065915.6-560108 which are discussed in
  the notes of Table~\ref{tab:slist}. 
\end{enumerate}
The two methods of estimating the false detections agree well. We note
that these calculations only gives a statistical measure of the number
of spurious sources, and not \emph{which} those sources are. However,
it is more likely that the least significant sources are false
detections.

\section{Cluster lens models and number counts}
\label{sec:cluster-lens-models}

\begin{table*}[t!]
  \centering
  \caption{Details about the cluster mass models}
  \begin{tabular}[t]{l c c c c c c l}
    \hline
    \hline
    Cluster name & Redshift & Mass & Scale radius &
    $c$ & Rel. pos. & Reference \\
    &  & [$ 10^{14} M_{\odot}$] & [arcmin] &  & [arcsec] &     \\
    \hline
    \vspace{1ex}
    Abell~2163 &  0.203 & $22$ & 4.4  & 3.6 & $0,0$ & \cite{RadovichPuddu:2008aa}  \\
    Bullet Cluster &  0.296 &  $31$  & 4.0 & 3.3 &  $-86,-24$ & \cite{Clowe:2004rr} \\
    \vspace{1ex}
    & &  $8.0 $ & 0.7 & 5.3 & $86,24$ \\  
    Abell~2744 &  0.308 &  $11$   & 2.5  & 3.5 & $0,0$ & \cite{BoschinGirardi:2006aa}  \\
    \vspace{1ex}
    &  & $3.5$  & 1.5 & 3.9  & $48,135$  \\  
    AC~114 & 0.312 &  $12$  & 2.6 & 3.5  & $0,0$ & \cite{CampusanoPello:2001aa} \\
    &  & $4.3$  & 1.7  & 3.9 & $75,-75$ \\
    \vspace{1ex}
    &  & $2.3$  & 1.2  & 4.2  & $80,30$ \\
    MS~1054-03  &  0.823  & $3.4$  & 1.0 & 2.9  & $0,0$ & \cite{HoekstraFranx:2000aa}\\
    &  & $3.4$ & 1.0  & 2.9 & $50,25$ \\
    &  & $3.4$ & 1.0  & 2.9 & $-60,-20$ \\
    \hline
  \end{tabular}
 \label{tab:models}
\end{table*}

The clusters in our survey were partly chosen for their lensing
properties because their high masses lead to areas of high
magnification. In order to estimate the magnification of the detected
sources, knowledge of the mass distribution of the clusters is
required. The following calculations use the thin lens approximation.

The magnification factor due to a gravitational lens is given by the
relation
\begin{equation}\label{eq:mag}
\mu = \frac{1}{\det \mathcal{A}},
\end{equation}
where $\mathcal{A}$ is the Jacobian of the lens equation,
\begin{equation}\label{eq:jacobian}
  \mathcal{A}(\boldsymbol{\theta})=\frac{\partial \boldsymbol{\beta}}{\partial
    \boldsymbol{\theta}}=\left(\begin{array}{cc} 1-\kappa -\gamma _1 & -\gamma _2\\
      -\gamma _2  & 1-\kappa +\gamma _1
\end{array}\right).
\end{equation}
$\kappa$ is the convergence of the lens while $\gamma_1$ and
$\gamma_2$ are the components of the complex shear.

We model the clusters as a superposition of one or more spherically
symmetric Navarro, Frenk \& White (NFW) mass profiles
\citep{NavarroFrenk:1997aa}. For an NFW profile, the convergence is
\citep{TakadaJain:2003aa}

\begin{equation}
\kappa(\theta)=\frac{\Sigma (\theta)}{\Sigma _{\text{crit}}} =
\frac{M_{\text{vir}}fc^2}{2 \pi r^2 _{\text{vir}}
  \Sigma_{\text{crit}}} \, F(c\theta/\theta
_{\text{vir}}),
\end{equation}
and the shear is
\begin{equation}
\gamma (\theta) = \frac{M_{\text{vir}}fc^2}{2\pi
r^2_{\text{vir}} \Sigma
_{\text{crit}}} \, G(c\theta /\theta _{\text{vir}}),
\end{equation}
where the critical surface mass density
\begin{equation}
\Sigma _{\text{crit}} = \frac{c^2}{4\pi G}\frac{D_S}{D_L D_{LS}},
\end{equation}
where $D_L$ is the angular-diameter distance to the lens, $D_S$ the
distance to the source, and $D_{LS}$ the distance between the source
and the lens, and $F$, $G$ and $f$ are functions of the concentration
parameter which can be found in appendix B in
\cite{TakadaJain:2003aa}.

The virial radius of the mass distribution, $r_{\text{vir}}$ can be
calculated from the virial mass
\begin{equation}
r_{\text{vir}}=\left(\frac{3M_{\text{vir}}}{\rho _{\text{crit}}(z)4\pi \Delta
_c}  \right)^{1/3},
\end{equation}
where
\begin{equation}
  \frac{\rho _{\text{crit}}(z)} {\rho _{\text{crit}}(0)} =
  \frac{H^2(z)}{H^2_0}=\Omega_0 (1+z)^3+ \Omega
  _{\Lambda 0}
\end{equation}
is the critical density at the redshift of the lens in a flat
Universe. The virial overdensity $\Delta _c$ can be estimated from a fit
to numerical simulations \citep{Bryan:1998ve}:
\begin{equation}
\Delta _c = 18\pi ^2 +82x-39x^2,
\end{equation}
where $x=\omega_m (z) -1$ and $\omega _m (z) = \Omega_0 (1+z)^3H_0^2/H^2(z)$.

We estimate the concentration parameter $c$ for a certain mass and
redshift using a fit to X-ray luminous clusters of galaxies
  \cite{EttoriGastaldello:2010aa}
\begin{equation}\label{eq:conc}
c(M_{\text{vir}},z) =
\frac{10^A}{1+z}\left(\frac{M_{\text{vir}}}{M_*}\right)^{B},
\end{equation}
with $M_*=1.0\times 10^{15} M_{\odot}$ and the fitted parameters
$A=0.558\pm 0.008$ and $B=-0.451\pm 0.023$. The resulting
magnification factors and number counts do not depend strongly on the
assumed $c-M$ relation; using the relation from
\citet{BullockKolatt:2001aa} results in magnification factors that
differ from those derived from the \citet{EttoriGastaldello:2010aa}
fit at a level lower than the statistical uncertainties.

We wrote a computer program that generates magnification maps by
solving Eq.~(\ref{eq:mag}) on a two-dimensional grid. Each cluster was
modeled as one or a sum of NFW halos, whose masses were taken from
mass models in the literature. The parameters for these mass models
are summarized in Table~\ref{tab:models}. We do not include any
individual galaxies in our models. We assume the background sources
(the \emph{source plane}) are at a redshift of $z=2.5$ but find that
the magnifications are not particularly sensitive to changes in source
redshift\footnote{Setting the redshift of the source plane $z=2.0$ and
  $z=3.0$ results in changes in the magnification factors of less than
  10\% and on average a change of 2\%. This uncertainty is smaller
  than the absolute flux calibration uncertainty, or the instrumental
  noise level for each source. }. By creating magnification maps for
the cluster fields, the magnifications of the detected sources could
be read out from their position in the maps.

A short discussion of each of the five cluster models follows:

\begin{itemize}

\item \textit{Abell~2163}: We used the mass from a weak lensing analysis
  performed by \cite{RadovichPuddu:2008aa} and modeled the cluster as
  a single NFW profile. Work by \cite{MaurogordatoCappi:2008aa}
  suggests that the cluster is an ongoing merger and that the mass
  distribution is elongated, which we do not account for in our simple
  model.

\item
\textit{1E~0657-56} is the most massive cluster in the survey and gives the
largest area with high magnification of the five clusters. It consists
of two components, one main cluster and a smaller
subcluster. \cite{Clowe:2004rr} fitted the main cluster to a NFW
profile and measured the mass of the subcluster using aperture
densitometry \citep{CloweLuppino:2000ab,FahlmanKaiser:1994aa}.

\item
\textit{Abell~2744} is made up of two subclusters aligned along the
line-of-sight \citep{BoschinGirardi:2006aa}. The subclusters have a
mass ratio of 3:1 as estimated from a fit to NFW profiles by
\cite{BoschinGirardi:2006aa}. We use their mass estimate as masses for
two concentric NFW profiles.

\item
\textit{AC~114}: We use the results from \cite{CampusanoPello:2001aa} for
our model. They improved upon a previous lensing model by
\cite{NatarajanKneib:1998aa}. Their model of AC~114~is made out of a
central cluster component, two smaller subclusters and a galaxy-scale
component centered on each bright cluster galaxy. We include the three
large components but not the galaxy-scale components in our model.

\item
\textit{MS~1054-03} consists of three distinct mass concentrations. We
model the cluster as three NFW profiles with masses estimated from a fit
of three singular isothermal sphere profiles performed by
\cite{HoekstraFranx:2000aa}.
\end{itemize}

\begin{figure*}[h!]
  \centering
  \includegraphics[width=17cm]{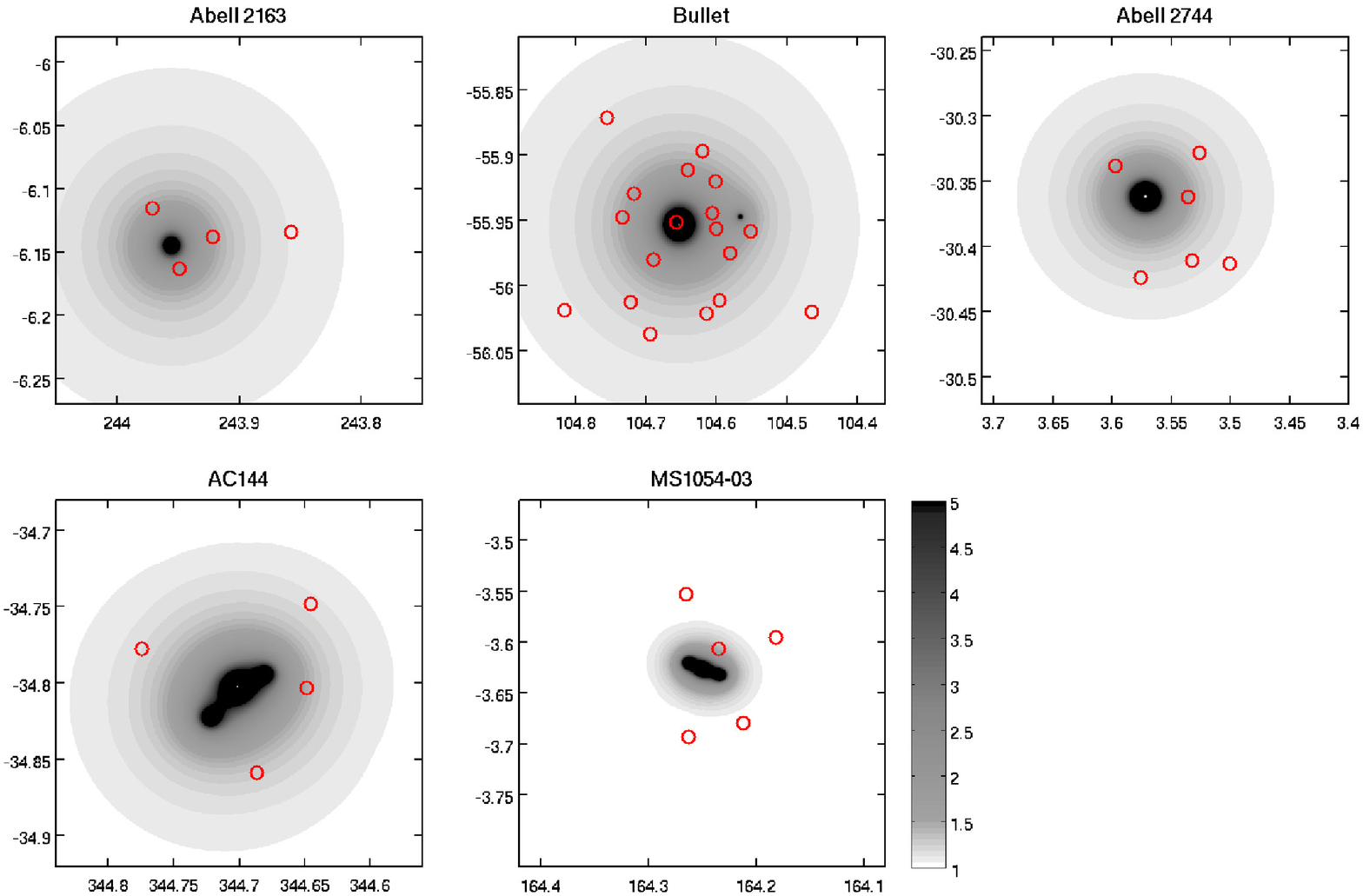}
  \caption{Magnification maps for the five clusters. The positions of
    the detected submm sources are marked on the maps with circles.}
  \label{fig:models}
\end{figure*}

\begin{figure}[h!]
  \centering
  \includegraphics[width=8cm]{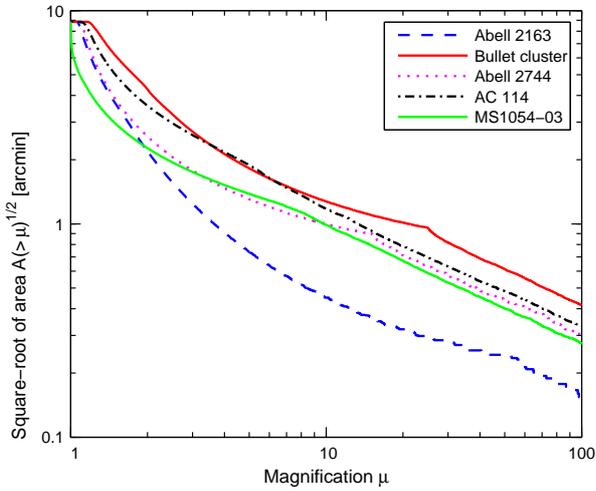}
  \caption{Square-root of the area with magnification $>\mu$ as
    function of magnification ($\mu$) for the five cluster models in
    the sample.}
  \label{fig:magar}
\end{figure}

\noindent
The resulting magnification maps are displayed in
Fig.~\ref{fig:models}. The positions of the detected sources for each
cluster are overlaid. The magnification values are listed in
Table~\ref{tab:slist}, as well as the demagnified flux densities. In
Fig.~\ref{fig:magar} we show the square root of the area (in the image
plane) that has a certain magnification factor or larger. This figure
shows the complex interplay between mass and redshift that determines
whether there are areas of high magnification. Abell~2163, which is the
second most massive cluster, is a less effective lens than AC~114~because
it is at lower redshift and therefore its mass is distributed over a
larger area on the sky.

\subsection{Number counts analysis}
\label{sec:number-counts}

Since gravitational lensing affects both the observed flux density of
a source and the area surveyed, we have to make corrections when
calculating number counts. The observed flux of the sources must be
demagnified to estimate the intrinsic flux. The intrinsic flux of a
source is related to the observed flux by $S_{obs}=\mu S_\text{i}$
where $\mu$ is the magnification of the source. Also, because the
magnification is not constant across the maps, neither is the
sensitivity. 

We consider the central 10$'$ of the maps where the noise is
approximately constant (see Fig.~\ref{fig:radrms} and the discussion
in Sect.~\ref{sec:survey-completeness}). We then impose the same
significance criterion as for the submm maps: that a source must have
a signal-to-noise ratio of 3.5 or higher in order to be reliably
detected. This signal-to-noise ratio corresponds to a certain minimum
observed flux density $S_{\text{min}}$.  A source with intrinsic flux
density of $S_\text{i}$ is then only detected if it lies in a region
with magnification $\mu \geq S_{\text{min}}/S_\text{i}$.

The area of this region in the lens plane, $A_{eff,l}(>S)$, is the
effective area that we are surveying for sources of a certain intrinsic
flux density or greater. The area in the lens plane corresponds to a
smaller area in the source plane due to magnification and the
effective area we are surveying in the source plane is
\begin{equation}
 A_{eff,s}(>S) = \sum _n A_{n} \mu _n,
\end{equation}
where $A_n$ is the area in the lens plane of a single area element and
$\mu _n$ is the magnification of that particular element. In our case
$A_n$ corresponds to the area of one pixel and $\mu _n$ the
magnification of that pixel. Thus a single detected source corresponds
to number count of $1/A_{eff,s}(>S)$ sources per unit area.

We also account for the effects of incompleteness in the maps, using
the Monte Carlo simulations described in
Sect.~\ref{sec:survey-completeness}. For each source we detect with a
flux density corresponding to a completeness of $C$ we expect there to
be on average $N_{\text{und}}=1/C-1$ undetected sources with the same
observed flux density.

By assuming that those undetected sources are uniformly distributed in
the map we can calculate the probability that they have a certain
intrinsic flux.  This probability is
\begin{equation}
P(S_{\text{int}}|S_{\text{obs}})=A_{\text{obs}\rightarrow\text{int}}/A_{\text{
field}}
\end{equation}
where $A_{\text{obs}\rightarrow\text{int}}$ is the area in the image
plane which has a magnification in the interval required to place a
source with an observed flux density $S_{\text{obs}}$ into the bin
corresponding to $S_{\text{int}}$. $A_{\mathrm{field}}$ is the total image
plane area of the field in which the source lies.

\subsection{Resulting number counts}
\label{sec:result-numb-counts}

\begin{figure}[h!]
  \centering
  \includegraphics[width=8cm]{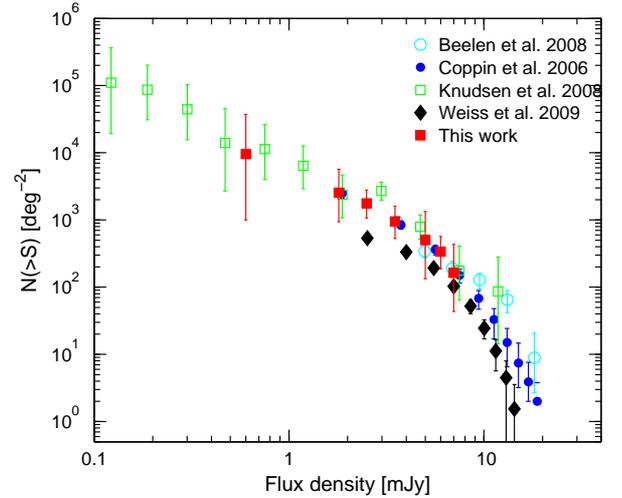}
  \caption{Derived number counts for the survey (filled boxes)
    compared with results from previous observations.}
  \label{fig:nc}
\end{figure}

The resulting number counts are shown in Fig~\ref{fig:nc}, together
with the results from the surveys of \cite{BeelenOmont:2008aa} and
\cite{WeisKovacs:2009ab} carried out with LABOCA~at 870~$\mu
\mathrm{m}$~as well as those from SCUBA surveys of
\cite{CoppinChapin:2006aa} and \cite{Knudsenvan-der-Werf:2008aa} at
850~$\mu \mathrm{m}$, for which the flux values were scaled from
850~$\mu \mathrm{m}$~to~870~$\mu \mathrm{m}$~with a spectral index of
2.7. There is generally good agreement with previous results. A much
larger survey of lensing clusters would be needed to reduce the
uncertainties and make more secure prediction about the dim submm
galaxies. Uncertainties for the number counts were calculated from
Poisson statistics, using the tables in \citet{Gehrels:1986aa}. The
number counts and their uncertainties are presented in
Table~\ref{tab:nc}.

\begin{table}[h!]
  \centering 
\caption{Number counts.}
  \begin{tabular}[h!]{ c l c}
    \hline
    \hline    
    {$S_{\mathrm{870 \, {\mu}m}}$} & $N(>S)$ & $N_{\mathrm{source}}$ \\
     $[\mathrm{mJy}]$ & deg$^{-2}$ &  \\
    \hline
    \vspace{1ex} 
    0.6 & $ 9570_{ 1005}^{37226} $ & 1 \\ 
    \vspace{1ex} 
    1.8 & $ 2552_{  937}^{ 5683} $ & 3 \\ 
    \vspace{1ex} 
    2.5 & $ 1771_{ 1069}^{ 2797} $ & 9 \\ 
    \vspace{1ex} 
    3.5 & $  951_{  529}^{ 1599} $ & 7 \\ 
    \vspace{1ex} 
    5.0 & $  501_{  133}^{ 1334} $ & 2 \\ 
    \vspace{1ex} 
    6.0 & $  337_{  188}^{  567} $ & 7 \\ 
    \vspace{1ex} 
    7.0 & $  164_{   44}^{  435} $ & 2 \\ 
    \hline
  \end{tabular}
  \label{tab:nc}
\end{table}

\section{Stacking analysis}
\label{sec:stacking-analysis}

\begin{figure*}[h!]
  \centering
  \includegraphics[width=18cm]{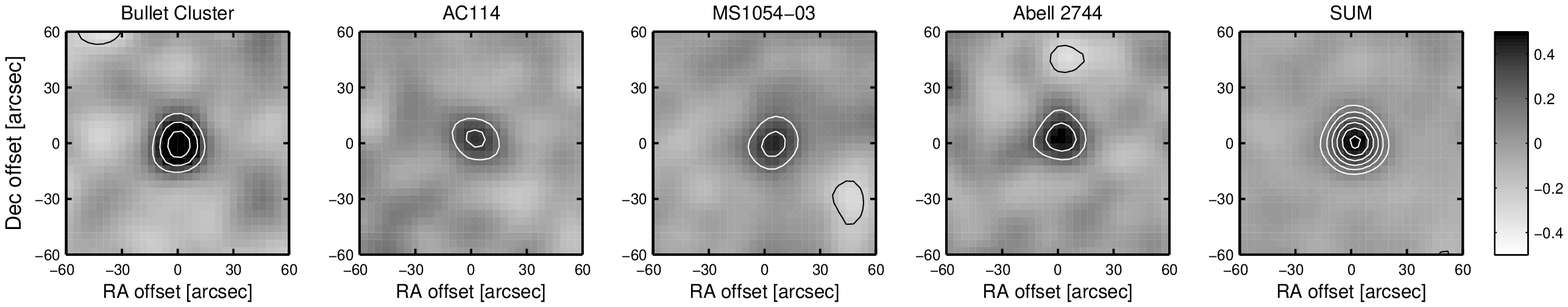}
  \caption{Stacked 870~$\mu \mathrm{m}$~maps (in units of
    $\mathrm{mJy\, beam}^{-1}$) on the 24~$\mu \mathrm{m}$~positions
    for the cluster fields with MIPS observations, overlaid with
    signal-to-noise contours. The white contours range between
    3$\sigma$ and 13$\sigma$ with an increment of 2$\sigma$, while the
    black countours show the $-3\sigma$ level. The maps have not been
    corrected for gravitational magnification. The fifth map is the
    coadded signal of the four individual stacked maps, which yields a
    $14.5\sigma$-detection. In Table~\ref{tab:stack} we present fits
    to the stacked maps.}
  \label{fig:stack}
\end{figure*}

We now turn to the analysis of the undetected sources in the maps. It
is well known that the most numerous contribution to the submm galaxy
population comes from dim sources with low flux densities (see for example
the number counts in Fig.~\ref{fig:nc}). Those dim sources cannot be
identified individually in the submm maps, but their presence can be
inferred statistically.

Stacking (coadding different parts of a map to lower the noise) has
proven to be an efficient method to reveal the underlying dim
population of submm sources. Such analysis has been performed by
e.g. \cite{ScottAustermann:2008aa,DunneIvison:2009aa,GreveWeibeta:2010aa}. Except
for Abell~2163, all fields in our sample have available Spitzer MIPS
24~$\mu \mathrm{m}$~data, as summarized in
Sect.~\ref{sec:spitzer-mips-data}.

The MIPS catalogs give us positions and flux densities of all 24~$\mu
\mathrm{m}$~sources in the fields, which we use to stack data in our
LABOCA~maps. We excluded MIPS positions which are farther than
6\arcmin~away from the LABOCA~map center, where the noise level is
rapidly increasing (see Fig.~\ref{fig:radrms}). Including positions
further out in the map would not lower the noise in the stacked
signal. We also excluded MIPS positions that lie closer than the size
of the LABOCA~beam from the cataloged submm sources in
Table~\ref{tab:slist}. For each 24~$\mu \mathrm{m}$~position we
extract submaps of size $2\arcmin \times 2 \arcmin$ from the
LABOCA~map (labelled $S_i$). We also extract the same region from the
noise map ($\sigma_i$), and use the following relation to stack the
submaps:
\begin{equation}
  \label{eq:1}
  S_{\mathrm{stack}} = \frac{ \sum\nolimits_i {S_i} / {\sigma_i^2}} {\sum\nolimits_i {1}/{\sigma_i^2}}, 
\end{equation}
i.e. a summation weighted by the variance. Lastly, we note that,
although no cataloged submm sources will enter the central position of
the stacked signal because those positions are discarded, they may
contaminate the outskirts of the stacked map. Therefore, we also
subtracted models of the cataloged submm sources from the LABOCA maps
before stacking. This lowers the noise levels in each of the stacked
maps but leaves the stacked flux densities unchanged.

In Fig.~\ref{fig:stack} we show the stacked images for the four
clusters. Each of the stacked maps shows a significant detection in
the central region, which is well fitted by a circular Gaussian of the
size of the beam. Flux densities and noise levels for the maps are
summarized in Table~\ref{tab:stack}. The noise levels of the stacked
maps are measured by subtracting the best-fit Gaussians and
calculating the pixel-to-pixel rms of the residual maps. The flux
densities of the stacked signals range from $\sim 350 -
820$~$\mu$Jy. This is equal to the mean observed flux density of the
sources that contribute to the stacked signal. In general terms, a
deeper MIPS catalog yields a lower 870~$\mu \mathrm{m}$~flux value
(comparing the stacked fluxes with the 24~$\mu \mathrm{m}$~depth the
deeper maps have a lower stacked signal). See also the following
section.

The stacked signal corresponds to an observed 870~$\mu
\mathrm{m}$~flux density. To investigate the intrinsic fluxes of the
dim galaxies we use the magnification maps derived from the cluster
models. We find the magnification factor for each 24~$\mu
\mathrm{m}$~position in the map and then stack the submaps extracted
from the LABOCA~map again, this time dividing each submap by the
magnification of the central source. This calculation is only valid
for the central part of the stacked map, since the magnification is
not constant across the submaps. We fit a circular Gaussian to the
stacked signals, and report the measurements in the fifth column of
Table \ref{tab:stack}. The demagnified stacked fluxes are lower than
the original.

\begin{table}[t!]
  \caption{Results from the stacking analysis. }
\centering
  \begin{tabular}[h!]{l c c c c}
    \hline
    \hline
    Cluster field & {$S_{\mathrm{870 \, {\mu}m}}$}\tablefootmark{a} &
    $\sigma$\tablefootmark{b} & $S/N$ & $S_{\mathrm{demag}}$\tablefootmark{c} \\
     &  [$\mu$Jy] & [$\mu \mathrm{Jy\, beam}^{-1}$]  & & [$\mu$Jy] \\
    \hline
   \multicolumn{5}{c}{Individual fields}\\
    Bullet Cluster & 815 &  88 & 9.3 & 591  \\
    AC~114 & 356 &  72 & 4.9  & 222 \\
    MS~1054-03 & 475 &  74 & 6.4  & 413 \\
    Abell~2744 & 514 &  82 & 6.2 & 345 \\
    \hline 
    \multicolumn{5}{c}{All fields}\\
    & 535 & 37 & 14.5 & 390 \\
    \hline   
  \end{tabular}
  \tablefoot{\tablefoottext{a} Measured by fitting a circular
    two-dimensional Gaussian of the size of the beam to the stacked
    signal. \tablefoottext{b} Pixel-to-pixel rms in the residual map where the
    best-fit Gaussian model was subtracted. \tablefoottext{c}
    Demagnified stacked signal taking the magnification at each
    24~$\mu \mathrm{m}$~position into account.}
 \label{tab:stack}
\end{table}

Finally, we coadd the four stacked signals, weighted by their
noise-maps, to find the total stacked 24~$\mu \mathrm{m}$~signal for
the entire survey (excluding Abell~2163). The map, shown in the fifth
panel in Fig.~\ref{fig:stack}, is a $14.5\sigma$--detection, as
reported in Table~\ref{tab:stack}. When performing the same operation
on the stacked maps corrected for gravitational magnification, we find
a mean signal of 390~$\mu$Jy for the four cluster fields. What would
be the properties of an submm galaxy with such a flux density?
Assuming a median redshift of $z=2.2$ \citep{ChapmanBlain:2005aa},
dust temperature $T_d \sim 40$~K and dust emissivity index $\alpha=2$,
the 870~$\mu \mathrm{m}$~flux density of 390~$\mu$Jy corresponds to a
far-infrared luminosity $L_{\mathrm{FIR}} \sim 6.4 \times 10^{11} \,
L_{\odot}$ \citep[eq. 11 in][]{De-BreuckNeri:2003aa}. By assuming that
the submm emission originates mainly from starburst phenomena, which
follows a Salpeter initial mass function with a low-mass cutoff
$m_l=1.6 \, M_{\odot}$, we can estimate a star-formation rate
$\mathrm{SFR} \sim 60\, M_{\odot} \, \mathrm{yr}^{-1}$ using Eq. 4 of
\cite{OmontCox:2001aa}. It is clear that the stacking analysis
uncovers a population of submm sources different from that detected
directly in the maps. The derived far-infrared luminosity and
star-formation rate depend on several assumptions about the underlying
submm population.

Several groups have obtained similar results when stacking on MIPS or
radio source positions. \cite{ScottAustermann:2008aa} found a stacked
signal of $324 \pm 25$~$\mu$Jy on $\sim 2000$ MIPS positions in the
AzTEC study of the COSMOS field \citep{ScovilleAussel:2007aa}. Their
MIPS map had a similar depth to those in the present study. Accounting
for the wavelength difference by scaling the AzTEC 1.1 mm flux density
with a mm/submm spectral index of $2-3$, the 870~$\mu \mathrm{m}$~flux
density would be $\sim 520 - 660$~$\mu$Jy. This is slightly higher
than the intrinsic LABOCA~flux density,
390~$\mu$Jy. \cite{GreveWeibeta:2010aa} found similar stacked flux
values in their study of the Extended Chandra Deep Field South.

\subsection{On the validity of the stacking detections}
\label{sec:valid-stack-detect}

In order to assess the validity of the detection, we perform the
stacking analysis on random positions in the LABOCA map, drawn from a
uniform distribution within the same map area as the 24~$\mu
\mathrm{m}$~maps. We run 20 such simulations for each field. None of
the simulated maps has stacked signal with a significant 870~$\mu
\mathrm{m}$~source in the center. The nondetection in the simulated
maps gives confidence in the stacking results on the real 24~$\mu
\mathrm{m}$~position, and shows a correlation between the MIPS and
LABOCA maps. We discuss the nature of this correlation in more detail
in the next section.

\begin{figure}[h!]
  \centering
  \includegraphics[width=8cm]{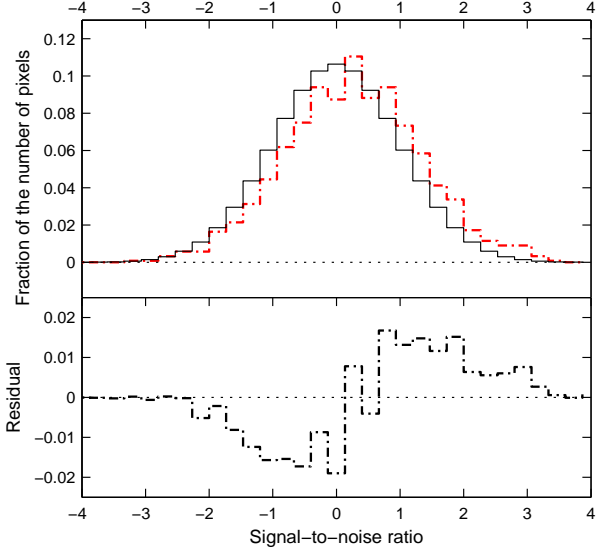}
  \caption{\emph{Top panel}: normalized histogram of the
    signal-to-noise values in the LABOCA~maps at the 24~$\mu
    \mathrm{m}$~positions (dash-dotted line), compared with the mean
    histogram from 20 sets of random positions (solid
    line). \emph{Lower panel}: Difference signal between the
    signal-to-noise histogram above, showing that the main part of the
    stacked signal is due to sources with signal-to-noise ratios
    between 0.5 and 2. This histogram clearly shows a deficit of
    points at negative signal-to-noise units and an excess at positive
    signal-to-noise units.}
  \label{fig:sthist}
\end{figure}

To investigate the significance of the sources that contribute to the
stacked signals, we extract pixel values at the 24~$\mu \mathrm{m}$~positions
from the signal-to-noise maps and compare them to the randomly
distributed positions. The histograms for all the 24~$\mu \mathrm{m}$~positions
not ascociated with a significant LABOCA~source is shown in
Fig.~\ref{fig:sthist}. It is compared with the histogram for randomly
distributed points, which has the shape of a Gaussian. The difference
signal between the two curves indicates that LABOCA~points with
significance $0.5\sigma < \mathrm{S/N} < 2\sigma$ contribute the most to the
stacked signal.

\subsection{The $S_{\mathrm{24 \, {\mu}m}}$ -- {$S_{\mathrm{870 \,
        {\mu}m}}$} relation}
\label{sec:which-sourc-contr}

\begin{figure}[h]
  \centering
 \includegraphics[width=8cm]{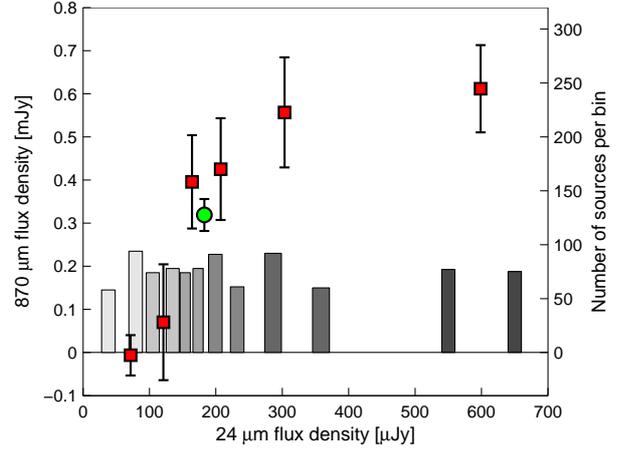}
 \caption{Results from the stacking analysis in MS~1054-03, AC~114~and
   Abell~2744, as described in
   Sect.~\ref{sec:which-sourc-contr}. \emph{Boxes}: measured flux
   density in the maps when dividing the 24~$\mu \mathrm{m}$~positions
   into six equal parts, and stacking them. The stacked flux densities
   are measured by fitting a circular Gaussian to the maps, except in
   the lowest two $24$~$\mu \mathrm{m}$~flux bins where no significant
   stacked signal was detected, and we instead measured the flux
   density in a circular aperture with the diameter of the beam
   FWHM. \emph{Circle}: the stacked signal for all the three
   clusters. \emph{Shaded bars}: histograms indicating the
   distribution of 24~$\mu \mathrm{m}$~flux densities within each of
   the six flux bins. Each of the bins have two bars and the different
   shades of gray discriminate between them. In the highest flux bin
   the rightmost bar includes all the 24~$\mu \mathrm{m}$~sources with
   flux densities larger than 650~$\mu$Jy. All flux values were
   corrected for gravitational magnification.}
  \label{fig:histstack}
\end{figure}

So far, we have investigated the signal resulting by stacking
LABOCA~sub-map at each MIPS position within the fields. However, it is
not plausible that all MIPS sources contribute equally to the stacked
map. By choosing subsets of the total MIPS catalog with different
24~$\mu \mathrm{m}$~flux values, we examine a possible correlation
between the flux density of the stacked signal and the 24~$\mu
\mathrm{m}$~flux density. We perform this analysis in the fields of
MS~1054-03, AC~114~and Abell~2744. There are 918 24~$\mu
\mathrm{m}$~sources in the three fields. We divide the catalog into
six sub-catalogs with equal number of sources (138), with median
24~$\mu \mathrm{m}$~flux densities of 72, 121, 164, 207, 303 and
599~$\mu$Jy, and perform the stacking analysis for each of them. Both
the 24 and 870~$\mu \mathrm{m}$~fluxes are demagnified. We find
significant signals in the four highest 24~$\mu \mathrm{m}$~flux
bins. The results are plotted in Fig.~\ref{fig:histstack}. At low
24~$\mu \mathrm{m}$~flux ($S_{\mathrm{24 \, {\mu}m}}$$<300$~$\mu$Jy)
we find a linear relation between the 24~$\mu \mathrm{m}$~and the
870~$\mu \mathrm{m}$~flux. At higher flux densities a turnover occurs
and the curve flattens out. The results are along the same lines as
those of \cite{GreveWeibeta:2010aa} who find a flattening of the
$S_{\mathrm{24 \, {\mu}m}}$~--~{$S_{\mathrm{870 \, {\mu}m}}$}~relation
at $S_{\mathrm{24 \, {\mu}m}} \sim 350$~$\mu$Jy, in their stacking
analysis.

\cite{GreveWeibeta:2010aa} argue that the linear relation at low MIPS
fluxes is an indication that those sources are dominated by star
formation, whereas the flattening of the curve at larger
24~$\mu \mathrm{m}$~fluxes is due to contamination by active galactic nuclei
(AGN). The mid-IR flux is sensitive to warm dust, which is likely to
be heated by an AGN. The 870~$\mu \mathrm{m}$~flux is more sensitive to colder
dust, heated by starbursts. While the mid-IR flux increases the
870~$\mu \mathrm{m}$~flux stays constant, because it is not sensitive to the
warm dust emission.

\subsection{A MIPS source contributing to the stacked signal}
\label{sec:spitzer-comparison}

It is beyond the scope of this paper to present a full
multi-wavelength analysis and comparison between the MIPS and
LABOCA~maps. Here, we note one source which has a large magnification
and is almost detected in the LABOCA~map. \cite{RigbyMarcillac:2008aa}
presented Spitzer/IRS spectroscopy of lensed galaxies, and discussed
one source in the center of AC~114, gravitationally magnified by a
factor of 9.7, and at a redshift of $z=1.47$.  At this position in our
LABOCA~map there is a positive signal with a significance of $3.4
\sigma$ and flux density $\sim 4$~mJy.

\begin{figure}[h!]
  \centering
  \includegraphics[width=7cm]{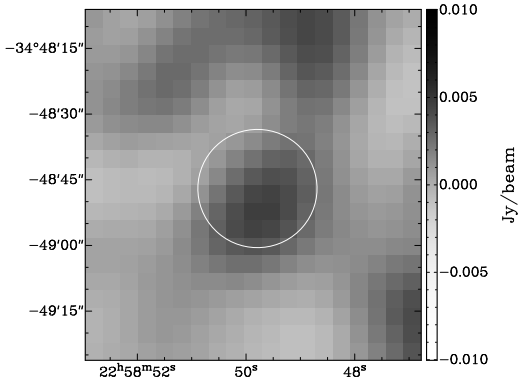}
  \caption{Postage stamp cutout of the core in AC~114. The circle
    (with diameter of the angular resolution of the LABOCA~map,
    $\mathrm{FWHM}=27.5\arcsec$) denotes the position of the
    Spitzer source at $z=1.47$, as reported by
    \cite{RigbyMarcillac:2008aa}.}
  \label{fig:irs}
\end{figure}

Figure~\ref{fig:irs} shows a postage-stamp cutout of the region around
the source. The LABOCA~counterpart to the Spitzer source is just below
our detection limit, and we would need more data to confirm it. If the
870~$\mu \mathrm{m}$~source is real, the high magnfication value would
mean that its intrinsic flux density {$S_{\mathrm{870 \,
      {\mu}m}}$}$<0.5$~mJy. We note that our lensing model for
AC~114~gives a magnification value of $\mu=6$ for the Spitzer source
position. Given that it lies very close to the brightest cluster
galaxy in AC~114, the small discrepancy between the two magnification
values are likely due to the lack of modeling of individual cluster
member galaxies in this work.

\section{Contribution to the extragalactic background light}
\label{sec:contribution-ebl}

We have described and quantified the submm emission from two types of
objects in the paper: the significantly detected sources in the
maps, and those whose mean flux density is inferred from the stacking
analysis. We now estimate the fraction of the Extragalactic
Background Light (EBL) detected in the far-infrared by COBE
\citep{PugetAbergel:1996aa,FixsenDwek:1998aa,DwekArendt:1998aa} that
our survey has resolved. At 870~$\mu \mathrm{m}$~the surface brightness of the
EBL is $44 \pm 15$~Jy~deg$^{-2}$ \citep{GreveWeibeta:2010aa}.

The total flux density of the detected submm sources that lie within
the 10 arcminute central region is $\sim 250$~mJy. This corresponds to
a surface brightness of 2.3~Jy~deg$^{-2}$, thus showing that the
LABOCA~observations have resolved $\sim 5\%$ of the EBL into significant
$>3.5\sigma$-sources. Another way to calculate the contribution to the
EBL is to use the number counts, which have been corrected for
completeness. Integrating the number counts yields a surface
brightness of 5.7~Jy~deg$^{-2}$, corresponding to $\sim 13\%$ of the
EBL.

Turning to the stacked signal, where we detected a mean observed flux
density of $535 \pm 37$~$\mu$Jy at 1278 positions in an area of $\sim
500 \, \mathrm{arcmin}^2$, the surface brightness is $\sim
4.9$~Jy~deg$^{-2}$ and the stacked submm signal thus corresponds to
11\% of the EBL. Thus, in total our observations have uncovered the
source of 24\% of the EBL. Because gravitational lensing preserves
surface brightness we can choose to perform this calculation using
either the observed or intrinsic flux densities and areas.

Other authors find different EBL
contributions. \cite{Knudsenvan-der-Werf:2008aa} find that their
observations resolve almost all of the extragalactic background light,
since they discovered seven galaxies with sub-mJy intrinsic flux
levels, and thus probe the number counts very
deep. \cite{GreveWeibeta:2010aa} find that the contributions for the
stacked signal in the Extended Chandra Deep Field-South varies with
redshift between 10\% at $z=0.5$ to 40\% at $z=2$.

\section{Conclusions}
\label{sec:conclusions}

We used the LABOCA~receiver on APEX to carry out a submm survey of
five clusters of galaxies. The clusters act as gravitational lenses
and magnify background sources. The main results of the survey are
summarized below.
\begin{enumerate}
\item We discovered 37 submm sources, out of which 14 are new submm
  detections.  
\item We modeled the galaxy clusters as superpositions of spherical
  NFW halos and generated magnification maps for the five clusters.
\item The magnification maps were used to correct for the
  gravitational lensing and to obtain the intrinsic flux densities of 
  the detected sources.
\item We constructed number counts taking into account both the
  gravitational lensing and the varying completeness level. The number
  counts probe the sub-mJy level and are consistent with previous work
  within the uncertainties.
\item We performed a stacking analysis in the LABOCA~maps on positions
  of detected 24~$\mu \mathrm{m}$~sources in the fields. The stacking
  yields $>4.9 \sigma$ detections in all fields with MIPS coverage,
  reaching noise levels below $100$~$\mu$Jy, more than an order of
  magnitude deeper than the individual maps.
\item By dividing the 24~$\mu \mathrm{m}$~catalog in MS~1054-03,
  AC~114~and Abell~2744~into six equal halves, we find a linear
  relation between $S_{\mathrm{24 \, {\mu}m}}$~and {$S_{\mathrm{870 \,
        {\mu}m}}$}~at low 24~$\mu \mathrm{m}$~fluxes, followed by a
  flattening of the relation at $S_{\mathrm{24 \, {\mu}m}} \sim
  300$~$\mu$Jy. This behavior can be explained if the low MIPS fluxes
  trace star formation while the higher values are dominated by AGN
  heating.
\item The observations reveal a total of $\sim$24\% of the infrared
  extragalactic background light, where $\sim$13\% comes from the
  significant submm sources and $\sim$11\% comes from the stacked
  signal.
\end{enumerate}

\begin{acknowledgements}
  We would like to thank the APEX staff for excellent support during
  the observations. We also thank Rodrigo Parra for his help during
  observations of Abell~2744~in June 2008. We thank Martin Sommer for
  supplying us with his data on Abell 2163 and Eiichi Egami for
  letting us use his catalog of MIPS source in the Bullet Cluster
  field. We would like to thank Kirsten Knudsen for helpful
  discussions. We acknowledge support from the Swedish research
  council (Vetenskapsr{\aa}det). We thank the referee for very useful
  comments.
  
  This work is based in part on observations made with the Spitzer
  Space Telescope, which is operated by the Jet Propulsion Laboratory,
  California Institute of Technology under a contract with NASA. This
  research has made use of the NASA/IPAC Extragalactic Database (NED)
  which is operated by the Jet Propulsion Laboratory, California
  Institute of Technology, under contract with the National
  Aeronautics and Space Administration. This research has also made
  use of NASA's Astrophysics Data System Bibliographic Services.

\end{acknowledgements}

\bibliographystyle{aa}
\bibliography{16138}

\appendix

\section{Comparison with other submm observations}
\label{sec:comp-with-other}

\subsubsection*{MS~1054-03}
\label{sec:ac}

Even though the galaxy clusters in our sample have been well studied
across the electromagnetic spectrum, submm maps have been published
for only one system, MS~1054-03~\citep[][hereafter
K08]{Knudsenvan-der-Werf:2008aa} using SCUBA. Subsets of those SCUBA
data were previously analyzed by
\citealt{ChapmanScott:2002aa,ZemcovBorys:2007aa,Knudsenvan-der-Werf:2005aa},
with mostly similar source catalogs. Differences in the SCUBA source
catalogs are discussed by K08. We will now compare our catalog for
MS~1054-03~with theirs.

K08 reach a noise level of $0.86 \, \mathrm{mJy\, beam}^{-1}$ in the
deepest part of the map, while their area-weighted noise level is
$1.49 \, \mathrm{mJy\, beam}^{-1}$. They detect nine significant
sources in their map. The noise level in the LABOCA~map in the central
part of the map is $1.3 \, \mathrm{mJy\, beam}^{-1}$ (see
Fig.~\ref{fig:radrms}), while the average noise level within the
central 10\arcmin~is 1.6$~\mathrm{mJy\, beam}^{-1}$. The SCUBA map
covers the cluster region (14.4~arcmin$^2$) while the usable map area
in the LABOCA~map is $\sim 150$~arcmin$^2$.

Only one of our detected sources lies in the area covered by K08. It
is extended with respect to the LABOCA~beam, and has an angular size
of $30 \arcsec \times 35\arcsec$. Its flux density is $9.8\pm
1.8$~mJy.  In the same area, K08 report three sources, separated by of
25.3\arcsec, 25.8\arcsec and 18.0\arcsec. LABOCA, with a coarser
resolution than SCUBA, causes the three sources to blend together. We
compared the measured LABOCA~source size with a simple model of the
three K08-sources, constructed as a sum of three LABOCA~beam shaped
Gaussians (angular FWHM 19.5\arcsec). We then smoothed this model
image with a Gaussian of angular FWHM of the size of the LABOCA beam,
and then fitted an elliptical Gaussian to the resulting map,
similarily to what was done in the real LABOCA maps. The fitted source
has an angular size of $33\arcsec \times 40\arcsec$. The slightly
larger angular size of the model compared to the observed
LABOCA~source can be explained with uncertainties in the fitted FWHMs
and in the SCUBA and LABOCA~positions. The sum of the flux density of
the three SCUBA sources, scaled from 850~$\mu \mathrm{m}$~to 870~$\mu
\mathrm{m}$~with a submm spectral index of 2.7, is $11.9 \pm
1.5$~mJy. This is within the $1\sigma$ uncertainty interval of the
LABOCA~flux measurement.

The other six sources detected by K08 have 850~$\mu \mathrm{m}$~flux densities
that, when extrapolated to 870~$\mu \mathrm{m}$, are too faint to be detected in
the LABOCA~map.

\subsubsection*{Abell~2163}
\label{sec:abb}

\cite{2009A&A...506..623N} presented the first LABOCA~map of a galaxy
cluster detected in the Sunyaev--Zeldovich~increment, Abell~2163. A
bright point source close to the cluster center, with a flux density
of $11.9\pm 1.9$~mJy, was noted, but not discussed. Using the same
data set as in \citeauthor{2009A&A...506..623N}, but filtering out
most of the extended SZ-signal, we detect the same point source in our
map, with a flux density of $8.9\pm2$~mJy.

\subsubsection*{1E~0657-56}
\label{sec:cl}

The brightest submm source in the Bullet Cluster has been thoroughly
discussed (see \citealt{JohanssonHorellou:2010ab} and references
therein). Submm and FIR observations by the SPIRE and PACS instruments
on the Herschel satellite show counterparts of LABOCA~sources
\citep{RexRawle:2010aa,Perez-GonzalezEgami:2010aa} as part of the
Herschel Lensing Survey \citep{EgamiRex:2010aa}.

\end{document}